\documentclass[twocolumn,showpacs,preprintnumbers,amsmath,amssymb,footinbib,prl]{revtex4}
\usepackage{epsfig}
\usepackage{graphicx}
\usepackage{color}

\def\unit#1{\ensuremath{\mathrm{~#1}}}
\def\unite#1{\ensuremath{\mathrm{~#1}}}

\def\TRamsey{\ensuremath{T_{\mathrm{Ramsey}}}}
\def\Rb{\ensuremath{^{87}\mathrm{Rb}}}
\def\Tacc{\ensuremath{T_\mathrm{acc}}}
\def\deltasep{\ensuremath{\delta_\mathrm{sep}}}

\def\poub#1{}

\def\vert#1{{\color{green}}}
\usepackage{amsmath}

\begin{document}

\title{Large Momentum Beamsplitter using Bloch Oscillations} 

\author{Pierre~Clad\'e}
\affiliation{Laboratoire Kastler Brossel, UPMC, Ecole Normale
Sup\'erieure, CNRS,  4 place Jussieu, 75252 Paris Cedex 05, France}
\author{Sa\"\i da~Guellati-Kh\'elifa}
\affiliation{Laboratoire Kastler Brossel, UPMC, Ecole Normale
Sup\'erieure, CNRS,  4 place Jussieu, 75252 Paris Cedex 05, France}
\author{Fran\c cois~Nez}
\affiliation{Laboratoire Kastler Brossel, UPMC, Ecole Normale
Sup\'erieure, CNRS,  4 place Jussieu, 75252 Paris Cedex 05, France}
\author{Fran\c cois~Biraben}
\affiliation{Laboratoire Kastler Brossel, UPMC, Ecole Normale
Sup\'erieure, CNRS,  4 place Jussieu, 75252 Paris Cedex 05, France}


\begin{abstract}
The sensitivity of an inertial sensor based on an atomic interfermometer is proportional to the velocity separation of atoms in the two arms of the interferometer. In this paper we describe how Bloch oscillations can be used to increase this separation and to create a large momentum transfer (LMT) beamsplitter. We experimentally demonstrate a separation of 10 recoil velocities. Light shifts during the acceleration introduce phase fluctuations which can reduce the contrast of the interferometer. We precisely calculate this effect and demonstrate that it can be significantly reduced by using a suitable combination of LMT pulses. We finally show that this method seems to be very promising to realize LMT beamsplitter with several 10s of recoil and a very good efficiency. 
\end{abstract}

\maketitle

Over the past two decades, the impressive advances on the control and the manipulation of atomic de~Broglie waves by coherent light pulses had led to the emergence of highly sensitive atomic interferometers\cite{Peters1999,Snadden1998,Fixler2007,Gustavson2000,Dimopoulos2007}. Atom interferometry is based on coherent splitting and recombination of atomic wave packets. The sensitivity of the atomic interferometer, which is proportional to the spatial separation of the wave packets, can be enhanced by increasing the interaction time. This time is limited in some experiments: this is the case for gravity measurement where the free fall of atoms is a limitation as well as recoil velocity measurement. In our recent measurement of the recoil velocity \cite{cadoret2008}, this time was limited by the size of the vacuum cell. Indeed atoms accelerated to velocities close to 1600 recoils ($\simeq 10 \unit{m/s}$ for \Rb\ atoms) move of several centimeter during the measurement. Furthermore, vibrations or phase noise of lasers which are more important for longer interaction time are also limitations for the duration of the measurement. 

The velocity splitting is usually achieved by using coherent momentum exchange between atoms and light (Raman or Bragg transition)\cite{Bergman1997}: with such a process atoms are prepared in a superposition of two states with a well control relative velocity. In the case of Raman transitions and first order Bragg transitions (the most widely used configurations), this velocity difference $\delta v$ is fixed to $2v_r$ where $v_r$ is the recoil velocity. To circumvent the limitation due to the time measurement, one can enhance the interferometer sensitivity by increasing the velocity separation between its two arms. Several attempts have been investigated to increase this separation \cite{Gupta,Giltner1995,Weitz,Denschlag,Miffre2005,muller:180405}. 


In this paper, we investigate a method to realize the momentum transfer using the so called "Bloch oscillation" (BO) method \cite{Wilkinson, Peik}. In this scheme, the large momentum transfer beam splitter (LMTBS) is obtained in a two steps process: the first one is a regular beamsplitter which transfers 2 recoil velocities while the second one is a large momentum transfer pulse (LMT pulse), that uses BO to coherently accelerate atoms of one of the two wave packets. When the beam splitter is used to recombine wave packets, the order of the two steps is reversed. This method was initially suggested in Ref.~\cite{Denschlag}, where a proof of principle experiment is described. 

Compare to other methods, BO allow to transfer a large number of recoils with a high efficiency. This method seems therefore to be the most suitable for high precision measurements. In this letter we focus on one of the main systematic effect which is due to level shift of atoms in the lattice and produces a large phase shift in the interferometer. This phase shift depends on the position of atoms in the laser beam and the motion of atoms induces fluctuations which reduce the contrast of the interferometer. It is especially large in our experiment in which, compared to Ref.~\cite{Denschlag}, we use a non degenerate gas. We calculate this effect and demonstrate that it can be significantly reduced by using a suitable combination of LMT pulses.

\begin{figure}
\begin{center}
\includegraphics[width = .99\linewidth]{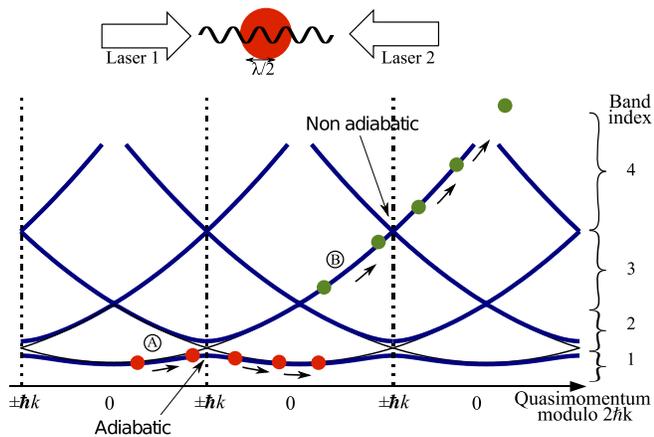}
\end{center}
\caption{Bande structure of the optical lattice. Trajectories of the accelerated (A) and non accelerated atoms (B).}
\label{fig:bande}
\end{figure}

The principle of Bloch oscillations \cite{Peik} consists in shining on the atoms two counterpropagating laser beams of similar frequencies ($\nu$ and $\nu + \delta\nu$). The interference of the two lasers results in an optical lattice of depth $U_0$ whose velocity is proportional to $\delta\nu$. When this lattice is accelerated, and under specific conditions, atoms are coherently accelerated. 
Because of the periodic potential, the eigenenergies of the system exhibit a band structure~\cite{Bloch}, each eigen state being described by a quasimomentum $q$ defined modulo $2\hbar k$ (where $k = 2\pi/\lambda$ is the laser wave vector) and a band index $n$. The figure~\ref{fig:bande} represents this band structure, where we have unfold the first Brillouin zone. There is a gap between bands either at the center ($q=0$) or at the edge of the first Brillouin zone ($q=\pm\hbar k$) when atoms are resonantly coupled. When an atom in the first band is subjected to a constant and uniform force, its quasimomentum increases lineary with time. If the force is weak enough to avoid adiabatic transition, the atom stays in the first band and therefore has a periodic motion (Bloch oscillations \cite{Zener}). The oscillations can occur also in higher band. However, the bandgap decreases with higher band index. Therefore, if an atom passes through the crossing at a given speed, the probability to make an adiabatic transition (to stay in the same band) will be higher for low value of the band index. 

The principle of the large momentum transfer beam splitter consists in creating a superposition of two wavepackets separated by 2 recoil velocities and in loading them in the optical lattice such that one wavepacket is in the first band (A, see Fig.~\ref{fig:bande}) and the second in the third band (B). A constant acceleration is then applied to the lattice. This acceleration (which acts like a force in the frame of the lattice) brings atoms to the edge of the first Brillouin zone. The acceleration of the lattice is small enough such that atoms in the first band have a large probability to make an adiabatic transition but high enough so that atoms in the third band change band. Each oscillation increases the momentum of the atom by $2\hbar k$. On the other hand, atoms in the third band (atoms that change band) are not accelerated.

\begin{figure}
\begin{center}
\includegraphics[width = .99\linewidth]{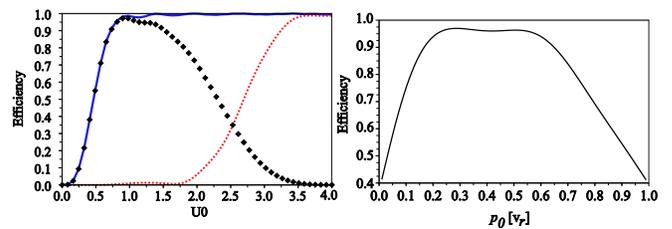}
\end{center}
\caption{Left: transfer probability as a function of the maximal optical depth $U_0$ of the lattice. $U_0$ is in unit of $8E_r$. The acceleration is of 2 recoils in 200\unit{\mu s}. Solid line (blue): transfer probability for the first band $\eta_{11}$; dashed line (red): for the third band ($\eta_{33} \approx 1-\eta_{34}$); diamond: Efficiency of the LMT pulse, $\eta =  \eta_{11}\eta_{34}$. Right: efficiency of the LMT pulse as a function of the initial momentum $p_0$.}
\label{fig:efficacite}
\end{figure}

For the efficiency of the LMT beamsplitter, the acceleration of the lattice should be weak enough to adiabatically accelerate atoms in the first band and strong enough to induce non-adiabatic transition for atoms in the third band. Reciprocally, for a given acceleration, the amplitude of the lattice should be in a given range. The figure~\ref{fig:efficacite} depicts the probability for an atom to stay in its band as a function of the lattice amplitude for the first (solid line) and third (dashed line) band.
There is clearly an intermediate regime where the probability $\eta_{11}$ for an atom to stay in the first band is high whereas the probability $\eta_{34}$ for an atom to leave the third band (and reach the fourth one) is also high. For an acceleration of 2 recoils in 200\unit{\mu s}, the total probability ($\eta =  \eta_{11}\eta_{34}$, diamond) presents a maximum around $U_0 = 8E_r$. The value of the maximum (97\%) depends on the duration of the acceleration and increases with this parameter.

We have plotted on the right side of Fig.~\ref{fig:efficacite} the total efficiency $\eta$ as a function of the initial momentum $p_0$. This efficiency is computed including the loading and unloading of atoms in the lattice: it is initially ramped up during a time $t_\mathrm{adiab}$, then accelerated during \Tacc\ and ramped down during  $t_\mathrm{adiab}$. By ramping adiabatically up and down the lattice, atoms from plane wave states are transferd to Bloch states and vice versa. As this process is not fully adiabatic, the efficiency of the LMT is reduced. 
At the center and the edge of the frist Brillouin zone ($q_0=0$ and $q_0=1$), the efficiency is strongly reduced because atoms cannot be loaded  adiabatically in the lattice (those points are initally degenerate). For the choosen parameters ($t_{\mathrm{adiab}}=150\unit{\mu s}$, $\Tacc=200\unit{\mu s}$, $N=2$ oscillations, $U_0 = 8 E_r$), we see that the efficiency is larger than 95\% on a large zone. An important issue is to maximize the width in initial momentum where the process is very efficient. Indeed atoms used in the interferometer have an initial velocity distribution selected by the Raman beam. The wider is the initial velocity distribution loaded into the LMT pulse, the higher is the number of atoms that contributes to the interferometer and so is the signal to noise ratio. We have optimized the efficiency by varying the amplitude and the temporal parameters keeping the total time $2t_{\mathrm{adiab}} + \Tacc = 500\unit{\mu s}$ constant.

\begin{figure}
\begin{center}
\includegraphics[width = .8\linewidth]{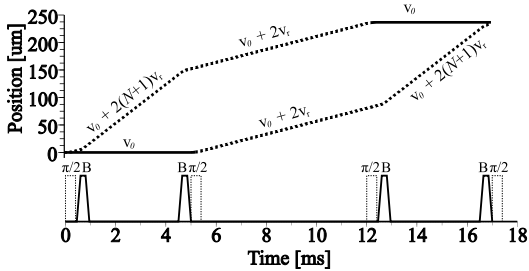}
\end{center}
\caption{Temporal sequence of the interferometer. Bottom: intensity of the Raman (dashed line) and Bloch (solid line) beam; Top: trajectories of atoms in the two arms; dashed and solid lines correspond to the two internal states of the atom.}
\label{fig:sequenceLMTSimple}
\end{figure}

One of the main drawback of the LMTBS based on Bloch oscillations is the light shift of atoms in the lattice. In the case of a blue detuned lattice, atoms in the first band are in a dark region and are almost not shifted while non-accelerated atoms in excited band see an average shift coresponding to the mean value of the potential of the lattice. For typical parameters, this light shift, much larger than $2\pi$, must be cancelled in order to run the interferometer. This cancellation occurs in the Mach-Zender configuration described on Fig.~\ref{fig:sequenceLMTSimple}. The configuration  used for the Raman pulses is similar to a regular interferometer with four $\pi/2$-pulses and the LMT pulses are added inside each pair of $\pi/2$ pulses used either for selection or measurement. With this scheme, the LMT pulses are applied symetrically on each arm of the interferometer, \textit{i.e.} one arm of the interferometer is initially in the first band and then in an excited band and \textit{vice versa} for the other arm. Therefore, the phase shift accumulated on each arm is the same and there is no systematic effect if the laser intensity seen by atoms is constant.  However, this is not the case because of temporal fluctuations of the laser intensity (leading to a phase noise in the interference pattern) or motion of atoms through the spatial profile of the laser beam (leading to a reduction of the fringes contrast). This last effect is not negligeable in our experiment: indeed with our experimental parameters, the typical relative variation of the laser intensity due to the motion of atoms in the gaussian profile of the laser is 6\unit{s^{-1}} (meaning a variation of $6\%$ in 10\unit{ms}). We numerically estimate that the sensitivity of the interferometer to such a linear temporal variation of the intensity is about 500\unit{rad.s}, leading to a typical phase shift of $\Delta \phi = 3\unit{rad}$. Consequelty, there is a complete supression of fringes. 

\begin{figure}
\begin{center}
\includegraphics[width = .8\linewidth]{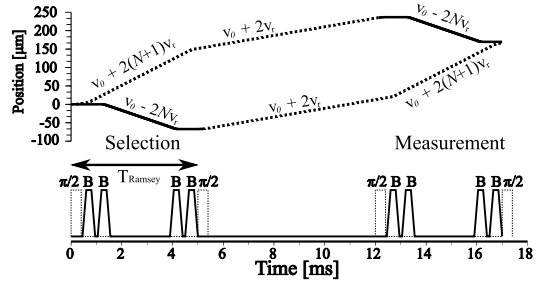}
\end{center}
\caption{Temporal sequence of the 4 LMT interferometer. Bottom: intensity of the Raman (dashed line) and Bloch (solid line) beam; Top: trajectories of atoms; dashed and solid line correspond to the two internal states of the atom.}
\label{fig:sequence}
\end{figure}

In order to run the inteferometer, we used a sequence (see Fig.~\ref{fig:sequence}) with 8 LMT pulses instead of 4. After (or before) each $\pi/2$ pulses,  we accelerate successively each arm of the interferometer with two LMT pulses of opposite directions. With this scheme, the second LMT pulse compensates about $90\%$ of the light shift~\footnote{The second LMT pulse do not fully compensate the phase shift of the first one, because during the second LMT pulse, the two arms of the interferometer are initially separated by more than 2 recoils and therefore are loaded in different bands compared to the first LMT pulse}. The sensitivity of the interferometer to a variation of the intensity is then reduced by a factor of 10 compared to the 4 LMT pulses interferometer. The resulting typical phase shift is $\Delta \phi = 0.3\unit{rad}$, which does not affect significantly the contrast of fringes. 
In the experiment, there is a correlation between the velocity and the position of atoms due to the time of flight before the begining of the interferometer. This results in a biased variation of the intensity which leads to a systematic effect in the interferometer. This effect can be canceled by inverting the order of the two LMT pulses for each beamsplitter. 

The experimental setup is similar to the one described in \cite{clade:052109}. We use \Rb\ atoms cooled using a magneto-optical trap followed by an optical molasses. The Raman beams, in a vertical configuration, are produced by two phase locked laser diodes and an acousto optic modulator (AOM) is used to control the shape of the $\pi/2$ pulses. The Raman beams are brought to the science chamber using an optical fiber. In order to reduce phase noise, a single polarization maintaining fiber is used. One of the two beams, selected by its polarization, is retroreflected in order to realize a counterpropagating transition. 

The Bloch beams are produced by a Ti:Sa laser. A new scheme has been implemented in order to reduce the phase of the Bloch beams, which affect the interferometer. As for the Raman beams, a single fiber and a retroreflecting mirror are used to create the optical lattice. But in contrary to the Raman beams, both beams used for Bloch are retroreflected. By sending to the AOM two frequencies ($\nu$ and $\nu +\delta\nu$) we create four lattices, one at the velocity of the atoms and three other lattices (one with the opposite velocity and two at rest).  In the experiment, we wait about 10\unit{ms} between the end of the molasses and the beginning of the interferometer. Because of the gravity, the velocity of the atoms is then about 20 recoils and the three other lattices are sufficiently out of resonance not to disturb the interferometer. The amplitude and frequencies sent to the AOM are controled using an RF chain based on an arbitrary waveform generator (similar to the one described in \cite{clade:052109}). 
In order to measure the velocity change of atoms between the first part and second part of the interferometer, the frequency of both the Raman and Bloch beams are changed in an identical way. The frequency of the central peak corresponds then to the Doppler effect induced by the variation of velocity. 


\begin{figure}
\begin{center}
\includegraphics[width = .99\linewidth]{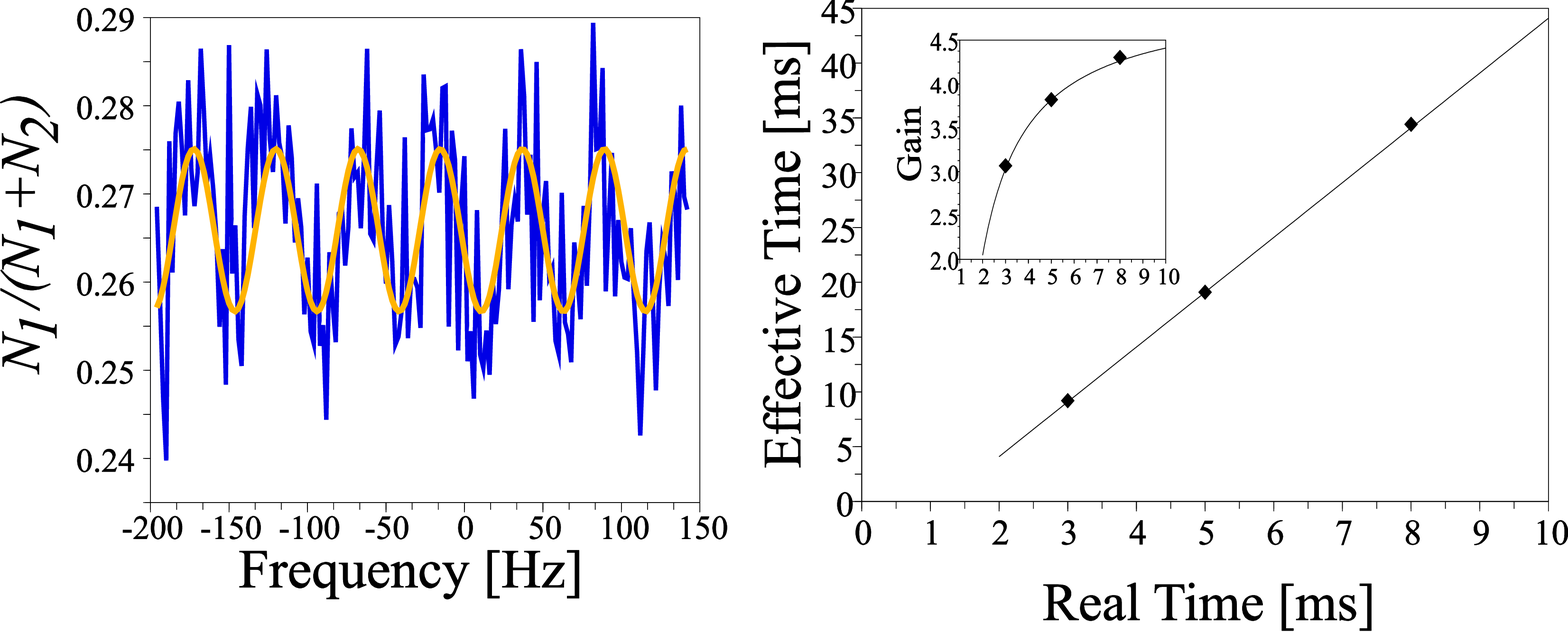}
\end{center}
\caption{Left: Interferences fringes observed using the scheme depicted Fig.~\ref{fig:sequence}. Right: Effective time $T_\mathrm{eff}$ as a function of the delay \TRamsey\ between the two $\pi/2$-pulses; solid curve: numerical calculation; Diamond: experimental results}
\label{fig:ResultatManip}
\end{figure}

The left side of Fig.~\ref{fig:ResultatManip} shows the proportion of atoms detected in the internal state $F=1$ as a function of the frequency difference between selection and measurement (see the temporal sequence of Fig.~\ref{fig:sequence}). The fringes sepration is $\deltasep = 50\unit{Hz}$ for this sequence. It corresponds to an effective time $T_\mathrm{eff}= 1/\deltasep$ of about 20\unite{ms}. We can compare this periodicity with the periodicity of fringes of an interferometer without LMT Bloch pulses. With the delay \TRamsey\ of 5\unit{ms} between the begining of the two Ramsey pulses, the periodicity would be 200\unit{Hz}. There is a gain of about 4 due to the use of LMT pulses. 

On the right side of Fig.~\ref{fig:ResultatManip} we have plotted $T_\mathrm{eff}$ for three different durations of the selection: keeping the same $\pi/2$-Bloch-Bloch sequence, we change \TRamsey. In the inset of the figure, we have plotted the gain in resolution (ratio between the effective time and \TRamsey) as a function of \TRamsey. At the limit where \TRamsey\ is long, the duration of the LMT Bloch pulses can be neglected and the gain is exactly $2N+1$, i.e. 5 for our parameters. At shorter time, this gain is smaller. 
The predicted effective time is plotted on Fig~\ref{fig:ResultatManip} (solid line). The good adjustement between the predicted time and our experimental data is a strong indication that the oscillations observed are due to the LMT pulses. 

The amplitude of the signal is roughly 10 times smaller than the amplitude without LMT pulses. Different sources seems to contributes to these losses of amplitude. Our model does not take into account the non resonnant lattices due to the retroreflecting configuration. Also, as explained earlier in the paper (Fig.~\ref{fig:efficacite}), we have to take into acount the initial distribution in quasi-momentum. We calculate that the total efficiency for 8 pulses (in the optimal configuration) is about 50\%. This is the maximal efficiency and higher losses are actually expected because the intensity of the lattice is not well controlled and also is not constant over the atomic cloud. Spontaneous emission, which for our parameters is of the order of 4\% for each of the LMT pulse, leads also to an additional reduction of 30\% of the signal. The use a laser beam with a wider waist, a higher intensity (and a larger detuning) should allow us to reduce significantly those effects. 

In order to further increase the velocity separation, one can imaging to have a two step process: the first step is identical to the one described earlier, with two Bloch oscillations. For the second step, we can increase adiabatically the depth of the lattice and then use a larger acceleration. A long time (and a relatively weaker lattice) is needed in order to realize the first separation with a good efficiency. However, for atom in the higher band, the Landau-Zener tunneling is larger and one can have a good efficiency with a shorter time and a deeper lattice. 
With an initial separation obtained with 2 oscillations in 250\unite{\mu s} with $U_0 = 6E_r$, a second separation with 8 oscillations in 150\unite{\mu s} (with $U_0 = 24 E_r$) and a total duration (taking into acount the adiabatic ramps) of 1\unit{ms}, the calculated efficiency is close to the 4 recoils LMT pulse (the losses are mainly due to the initial splitting). This scheme could then be used on both arms of the interferometer to realize a separation of 42 recoils.


In this paper, we have theoretically and experimentaly studied the implementation a large momentum beamsplitter in an atomic interferometer. It is possible to realize a separation of 4 recoils between the 2 arms of the interfermeter within 500\unite{\mu s}. The main issue to take care of is the light shift which degrades strongly the contrast of the interferometer. Our proposal allows us to reduce this effect by a factor of 10 and finally permits us to realize an interferometer with a separation of 10 recoils between the two arms. A gain of 4 in the resolution compared to a usual interferometer has been observed. This method seems to be very promizing for the realization of a beam splitter with separation of several tens of recoil velocity. The effect due to light shifts has been reduced by accelerating successivelly both arm of the interferometer. A way to suppress it systematically would be to accelerate both arms on the interfereometer simultaneously (instead of doing it successively) by applying two counter-propagating accelerated lattices. In this case light shift would be completely supressed. In the futur, we plan to combine this kind of LMT beam splitter with Bloch oscillations between the two pairs of $\pi/2$ pulses \cite{cadoret2008} to realize a new measurement of the recoil velocity and therefore improve the determination of the fine structure constant. 


\end{document}